\documentclass[prl, twocolumn,showpacs,preprintnumbers,amsmath,amssymb ]{revtex4}

\usepackage{graphicx}
\usepackage{dcolumn}
\usepackage{bm}

\begin{document}

\input epsf.sty



\title{On the long-time behavior of spin echo and its relation
to free induction decay}

\author{B. V. Fine}

\affiliation{
Max Plank Institute for the Physics of Complex Systems,
Noethnitzer Str. 38, D-01187, Dresden, Germany}

\date{November 12, 2004}

\begin{abstract}
It is predicted that (i) spin echoes have two kinds of generic long-time
decays: either simple exponential, or a superposition
of a monotonic and an oscillatory exponential decays; and (ii)
the long-time behavior of spin echo and the long-time behavior
of the corresponding homogeneous free induction decay are 
characterized by the same time constants. This prediction
extends to various echo problems both within and beyond
nuclear magnetic resonance. 
Experimental confirmation of this prediction
would also support the notion of the eigenvalues of time 
evolution operators in large quantum systems.
\end{abstract}
\pacs{76.60.Lz, 76.60.Es, 05.45.Gg, 03.65.Yz}


\maketitle


Calculations of spin echo (SE) envelope observed by
nuclear magnetic resonance (NMR) in solids  
are, in general, very difficult to do and, in practice,
mostly limited to simple sudden-jump models (e.g.,\cite{Klauder-etal-62}), 
which cannot capture complex 
non-Markovian dynamics induced by spin-spin interactions.
The goal of this work is to approach the SE problem from a new direction
by making the following two propositions:

1. The generic long-time behavior of SE envelope $S(2\tau)$ obtained
from the sequence 
[{\it $\pi/2$-pulse --- time $\tau$ --- $\pi$-pulse  ---
time $\tau$ --- detection}] has form either 
\begin{equation}
S(2 \tau) = C e^{- 2 \gamma  \tau}
\label{Smono}
\end{equation}
or
\begin{equation}
S(2 \tau ) = e^{- 2 \gamma \tau} 
[C_1 + C_2  \ \hbox{cos}(2 \omega \tau + \phi_S)],
\label{Sosc}
\end{equation}
where $C$, $C_1$, $C_2$, $\gamma$, $\omega$ and $\phi_S$ are some constants.

2. The constants $\gamma$ and $\omega$ describing the long-time SE decay  
are the same as the
long-time constants of the corresponding ``homogeneous'' 
free induction decay (FID) defined by the pulse sequence
[{\it $\pi/2$-pulse --- time $t$ ---  detection}]
and exhibiting generic long-time behavior\cite{Fine-04} of the form either
\begin{equation}
F(t) = A  e^{-\gamma t}
\label{Fmono}
\end{equation}
--- counterpart of Eq.(\ref{Smono}),  or
\begin{equation}
F(t) = A  e^{-\gamma t} \hbox{cos}(\omega t + \phi_F).
\label{Fosc}
\end{equation}
--- counterpart of Eq.(\ref{Sosc}). Here $A$ and $\phi_F$ are some additional
constants.
(The meaning of the ``homogeneous'' FID will be 
specified below.)

An important consequence of the above statement is that 
the long-time constants ($\gamma$ and $\omega$) of SE envelope 
can be obtained
from a much simpler FID calculation. At present,
a number of approximations exist (see, e.g., \cite{Fine-97,Fine-00}
and references therein), which can be used to
calculate the FID long-time constants with reasonable
accuracy.

In the following, I shall first
consider the case of spin-spin interaction only, and then
generalize the treatment to the presence of spin-lattice interaction
and to other echo problems both within and beyond NMR.

The simplest setting, which exposes the generic difficulties of the SE 
problem, involves two spin species $I$ and $S$ characterized by 
different Larmor frequencies. The relevant Hamiltonian is
$
{\cal H} = {\cal H}_{inh} +{\cal H}_{hom},
$
where ${\cal H}_{inh}$ contains the interaction of spins with
a slowly varying in space external field; and
${\cal H}_{hom}$ includes the spin-spin interaction and, sometimes,
the interaction with microscopically inhomogeneous external fields. 

The SE technique cancels the effect of ${\cal H}_{inh}$, 
so that the SE envelope
measured on $I$-spins, is given by:
\begin{eqnarray}
\nonumber
S(2 \tau) &= \hbox{Tr} &
\left\{  
e^{i {\cal H}_{hom} \tau} X^{\dagger}(\pi) e^{i {\cal H}_{hom} \tau}
\sum_i I_{ix} 
\right.
\\
&&
\times  
\left. e^{-i {\cal H}_{hom} \tau} X(\pi) e^{-i {\cal H}_{hom} \tau}
\sum_i I_{ix}
\right\},
\label{S}
\end{eqnarray}
where $I_{ix}$ is the $x$-component of the $i$th $I$-spin,
and $X(\pi)$ is the operator of the $\pi$-pulse acting on $I$-spins.

\begin{figure} \setlength{\unitlength}{0.1cm}

\begin{picture}(100, 48) 
{ 
\put(0, -2){ \epsfxsize= 3.1in \epsfbox{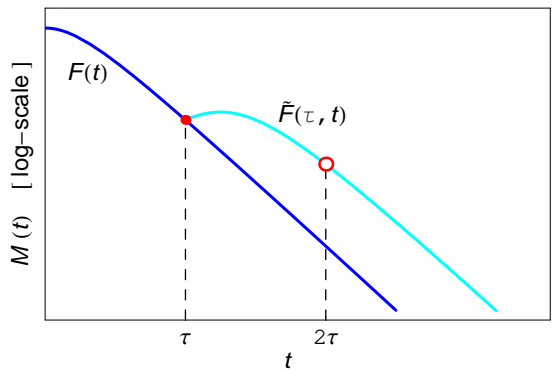} }
}
\end{picture} 
\caption{An example of the process, which generates a single point 
(open circle) of the spin echo envelope [see Eq.(\ref{M})].    
} 
\label{fig-M} 
\end{figure}

The process, which, according to Eq.(\ref{S}), generates 
a point of the SE envelope can be expressed as
follows:
\begin{equation}
M(t) = \left\{
\begin{array}{l@{\quad:\quad}l}
F(t) & 0 \leq t \leq \tau 
\\ 
\tilde{F}(\tau, t) &  \tau < t ,
\end{array}
\right. 
\label{M}
\end{equation}
where $F(t)$ is the homogeneous FID, i.e.
\begin{equation}
F(t) = \hbox{Tr} \left\{ 
e^{i {\cal H}_{hom} t} 
\sum_i I_{ix}
e^{-i {\cal H}_{hom} t} 
\sum_i I_{ix}
\right\},
\label{F}
\end{equation}
and $\tilde{F}(\tau, t)$ is the ``after-pulse'' 
response of the system in the absence of inhomogeneous broadening
(see Fig.~\ref{fig-M}).


Derivation of Eqs.(\ref{Smono},\ref{Sosc}) involves two steps: 
{\it step~1} --- to show that the generic long-time behavior 
of FID 
is given by Eqs.(\ref{Fmono},\ref{Fosc}); 
and, {\it step~2} --- to show that
Eqs.(\ref{Fmono}) and (\ref{Fosc}) for FID imply
Eqs.(\ref{Smono}) and (\ref{Sosc}) for SE.

{\it Step 1} has actually been made in 
Refs.\cite{Fine-00,Fine-04}
for the problems involving only one kind of equivalent spins.
These references contain detailed arguments and 
also summarize quite substantial
experimental\cite{Engelsberg-etal-74} and 
numerical\cite{Fabricius-etal-97,Fine-03} evidence 
that the long-time behavior (\ref{Fmono},\ref{Fosc})
is generic for isolated classical and quantum spin systems. 
That argumentation
applies directly to the present case, 
when ${\cal H}_{hom}$ is invariant under lattice
translations. [The random placement of spins
will be discussed later.]

I now proceed to {\it step~2} for the monotonic case (\ref{Smono},\ref{Fmono}).
The FID having the long-time behavior (\ref{Fmono}) can be written as:
\begin{equation}
F(t) = \left\{
\begin{array}{l@{\quad:\quad}l}
\hbox{non-universal} & 0 \leq t \leq \tau_0 
\\ 
A  e^{-\gamma t} &  \tau_0 < t ,
\end{array}
\right. 
\label{Fmono1}
\end{equation}
where $\tau_0$ is the characteristic time beyond which the difference
between the actual FID and its exponential asymptotics becomes negligible.

One can now observe that an
exponential decay  appearing 
in an equilibrium correlation function [such as $F(t)$]
likely indicates
that, in the corresponding linear response non-equilibrium 
problem\cite{Fine-04}, the
leading correction to the equilibrium density matrix of any subsystem
of spins evolves as 
\begin{equation}
\varrho_0(\mathbf{x}) e^{-\gamma t},
\label{rho0}
\end{equation}
where 
$\mathbf{x}$ is the set
of variables  describing the density matrix, and $\varrho_0(\mathbf{x})$ is 
some function.
An analogous statement applies to the probability distribution
of classical spins.

In the case of classical hyperbolic chaotic systems, Eq.(\ref{rho0})
has a rigorous justification as it constitutes a manifestation of 
a Pollicott-Ruelle resonance\cite{Ruelle-86,Gaspard-98}.
Beyond this very limited class of systems,  that equation
amounts to a very plausible hypothesis. 
In Refs.\cite{Fine-00,Fine-04}, the case has been made for extending
the notion of Policott-Ruelle resonances to
large isolated systems of classical and 
quantum spins. The numerical evidence for quantum analogs 
of Policott-Ruelle resonances
has also been reported for kicked quantum top\cite{Manderfeld-etal-01} and 
kicked chain of spins 1/2~\cite{Prosen-02,Prosen-04}.

In the regime describable by Eq.(\ref{rho0}), the effect of identical
manipulation of the system ($\pi$-pulse, in this case) should have 
self-similar 
effect on the density matrix
up to the prefactor $\hbox{exp}(-\gamma t)$. Since the same exponent 
also appears in Eq.(\ref{Fmono1}) ,  
the after-pulse response for $\tau > \tau_0$
can be factorized as
\begin{equation}
\tilde{F}(\tau, t) = F(\tau) F_1(t-\tau),
\label{Ftilde1}
\end{equation}
where $F_1(t-\tau)$ is some function

One can further stipulate that $e^{-\gamma t}$
dominates the long-time behavior of $F(t)$, not because of the specific
initial conditions, but simply because 
(i) $\gamma$ is the eigenvalue of a given 
time evolution operator; and (ii) among the eigenvalues selected by
the symmetry of the initial conditions\cite{Fine-04}, 
$\gamma$ has the smallest real part. 
Therefore, the density matrix created after the $\pi$-pulse should eventually
(after some delay time $\tau_1 \sim \tau_0$) approach the same asymptotics,
which implies that
\begin{equation}
F_1(t-\tau) = \left\{
\begin{array}{l@{\quad:\quad}l}
        \hbox{non-universal} 
    	& \tau \leq t \leq  \tau + \tau_1 
\\ 
	B e^{-\gamma (t - \tau)} &  \tau + \tau_1 < t .
\end{array}
\right. 
\label{F1mono}
\end{equation}
Thus, when $\tau > \hbox{max}[\tau_0, \tau_1]$, 
Eqs.(\ref{Fmono1},\ref{Ftilde1},\ref{F1mono}) yield
\begin{equation}
S(2 \tau) \equiv \tilde{F}(\tau, 2 \tau) = F(\tau) F_1(\tau) 
= A B e^{- 2 \gamma \tau},
\label{S2}
\end{equation}
which is equivalent to Eq.(\ref{Smono}).

Now I repeat {\it step~2} starting from the 
oscillatory FID (\ref{Fosc}). 
In this case,
\begin{equation}
F(t) = \left\{
\begin{array}{l@{\quad:\quad}l}
\hbox{non-universal} & 0 \leq t \leq \tau_0 
\\ 
f(t)  +
f^*(t) &  \tau_0 < t ,
\end{array}
\right. 
\label{Fmono2}
\end{equation}
where
\begin{equation}
f(t) = a  e^{-(\gamma + i \omega) t},
\label{f}
\end{equation}
and $a$ is a complex number.

In the long-time regime, the density matrix  should be dominated 
by two complex conjugate
self-similar terms:
\begin{equation}
\varrho_0(\mathbf{x}) e^{-(\gamma + i \omega) t} +
\varrho_0^*(\mathbf{x}) e^{-(\gamma - i \omega) t},
\label{rho0osc}
\end{equation}
where $\varrho_0(\mathbf{x})$ is now a complex-valued function.
After the $\pi$-pulse, 
\begin{equation}
\tilde{F}(\tau, t) = f(\tau) f_1(t-\tau) + f^*(\tau) f_1^*(t-\tau),
\label{Ftilde2}
\end{equation}
where 
\begin{equation}
f_1(t-\tau) = \left\{
\begin{array}{l@{\quad:\quad}l}
\hbox{non-universal} & \tau \leq t \leq  \tau + \tau_1 
\\ 
&
\\
\begin{array}{l}
		b_1 e^{-(\gamma + i \omega) (t - \tau)} 
		\\
		+ b_2 e^{-(\gamma - i \omega) (t - \tau)}
	\end{array} 
	&  \tau + \tau_1 < t .
\end{array}
\right. 
\label{F1osc}
\end{equation}
Here, $b_1$ and $b_2$ are complex numbers, which do not need to 
be the complex conjugates of each other, 
because $\varrho_0(\mathbf{x})$ is not real.
Thus, when $\tau > \hbox{max}[\tau_0, \tau_1]$, 
Eqs.(\ref{Fmono2},\ref{f},\ref{Ftilde2},\ref{F1osc}) give
\begin{eqnarray}
\label{S2osc}
S(2 \tau) =& f(\tau) f_1(\tau) + f^*(\tau) f_1^*(\tau) = 
|a| e^{-2 \gamma \tau} \ \ \ \ \ \ \ \ 
\\
\nonumber
& \times
\left[ 
|b_1| \hbox{cos}(2 \omega \tau - \phi_a - \phi_{b1})
+ |b_2| \hbox{cos}(\phi_a + \phi_{b2})
\right],
\end{eqnarray}
where $\phi_a$, $\phi_{b1}$ and $\phi_{b2}$ are the phases of
$a$, $b_1$ and $b_2$, respectively. Equation (\ref{S2osc}) 
is equivalent to  Eq.(\ref{Sosc}).

Several comments and generalizations are now in order:

1. When a non-Hamiltonian (spin-lattice) component is added to 
the above problem, it should only strengthen the reliability 
of Eqs.(\ref{Fmono},\ref{Fosc}), 
because the Markovian 
processes explicitly imply time-irreversibility and generically lead
to the exponentially decaying FIDs.
The rest of the proof then remains unchanged.

2. The exact study of echo in the presence 
of a single Markovian fluctuator\cite{Galperin-etal-03}
fully supports the long-time asymptotics (\ref{Smono}-\ref{Fosc}).

3. As generic, I consider the case, 
when $h_0 \sim W_0$, where 
$h_0$ is the typical value of local fields in ${\cal H}_{hom}$, and
$W_0$ their characteristic fluctuation rate.
In this case, $\tau_0 \sim \tau_1 \sim W_0^{-1}$.
The ``generic'' deviation from the generic
case can be described as a situation, when the  delay times $\tau_0$
and $\tau_1$ are anomalously long. Such a situation
may occur, when $h_0 \gg W_0$, or when the local fields or
the fluctuation rates are broadly
distributed.

4. Of particular concern can be the case,
when the lattice sites are randomly partitioned between
spins $I$ and $S$ representing, e.g., two different nuclear isotopes.
In this case, different
$I$-spins are not equivalent to each other.
Extending the analysis of Ref.\cite{Fine-04}, one can argue 
that, for each sub-ensemble of equivalent $I$-spins
the long-time behavior of FID is still given by Eqs.(\ref{Fmono},\ref{Fosc}), 
and,
therefore, the resulting signal $F(t)$ is the superposition of a continuous
distribution of exponents (\ref{Fmono},\ref{Fosc}). 
For nuclear spins in solids,
the distribution of parameter
$\gamma$ should be limited from 
below, because every spin experiences a finite fluctuating local field.
Therefore, as a matter of principle, the long-time
asymptotics (\ref{Fmono},\ref{Fosc}) will, eventually, be reached.
However, in the worst case scenario, the distribution of 
$\gamma$ can be very broad,
and, therefore, it can take impractically long time before the 
asymptotics (\ref{Fmono},\ref{Fosc}) becomes
pronounced. 

In many NMR settings, two factors, 
work against the above worst case scenario. Namely:
(i) The long-range nature of magnetic dipolar interaction between
nuclei implies that each $I$-spin interacts with a sufficiently large
number of  $S$-spins and other $I$-spins. Therefore, 
different $I$-spins, which, in a strict sense, are not equivalent, experience
statistically equivalent environments, which implies that the distributions
of $\gamma$ and $\omega$ are narrowly peaked. (ii) $I$-spins
belonging to different sub-ensembles are still  coupled to each other,
which should make spin dynamics more ``homogeneous''.

It should be noted, however, that, the averaging of 
the oscillatory SE (\ref{Sosc}) even
over a narrow distribution of  $\omega$ will, eventually, suppress
the beats leaving one with the monotonic asymptotics (\ref{Smono}).

\begin{figure} \setlength{\unitlength}{0.1cm}
\begin{picture}(100, 42) 
{ 
\put(5, -2){ \epsfxsize= 2.7in \epsfbox{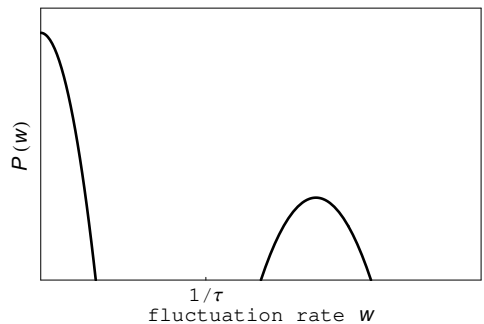} }
}
\end{picture} 
\caption{ Distribution of the local field fluctuation rates $P(W)$
consistent with the present theory. 
} 
\label{fig-distrib} 
\end{figure}

5. Beyond NMR, the language of free induction decays and echoes is frequently
applied to the situations, when there exists a distribution of local 
fields and their fluctuation rates.
The arguments presented in this work apply to the cases, when
the distribution of the fluctuation rates  
satisfy the condition shown in Fig.~\ref{fig-distrib}. 
Namely, the local fields can be divided
in two groups: very slowly fluctuating --- the effect of
these fields is cancelled by the SE pulse sequence; 
and those fluctuating with rates of 
the order of  $1/\tau_0$  or higher
--- they control both the [homogeneous] FID and the echo envelope.


6. Hu and Hartmann\cite{Hu-etal-74} have considered a mathematical model 
of infinitely dilute subsystem of $I$-spins surrounded 
by a large density of
randomly placed $S$-spins interacting with $I$-spins via 
magnetic dipolar interaction.
They have found 
that the long-time behavior of SE is given by 
$e^{-c \tau^{1/2}}$, where $c$ is a constant.
The difference between this asymptotics and the one given by 
Eq.(\ref{Smono}) is due to the fact that the above mathematical model
allowed for a non-zero probability that spins $S$ are absent in any
finite volume around spin $I$. 
Although the above probability decreases exponentially with the volume,
the corresponding fraction of $I$-spins produces an exponentially
slow contribution to the SE decay. The competition between these
two exponential factors then leads to a non-universal decay law.

\begin{figure} \setlength{\unitlength}{0.1cm}
\begin{picture}(100, 61) 
{ 
\put(0, -5){ \epsfxsize= 2.7in \epsfbox{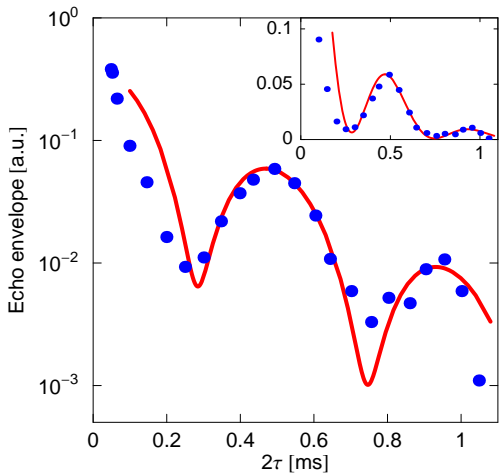} }
}
\end{picture} 
\caption{ 
Circles --- experimental SE envelope for $^{195}$Pt in Pt-Au alloy 
from Fig.~7 of Ref.\cite{Alloul-etal-67},
line ---  fit of the form (\ref{Sosc}). Inset: non-logarithmic
plot of the same data.
} 
\label{fig-alloul} 
\end{figure}

7. The experimental evidence for the monotonic echo 
asymptotics (\ref{Smono}) 
is easily available --- see, e.g.,  
Refs.\cite{Haase-etal-93,Pennington-etal-89}. 
The oscillatory SE decays are, presumably, more rare. 
Nevertheless, they  
have also been observed --- see, e.g., Refs.\cite{Alloul-etal-67,Wang-etal-84}. 
The data from Ref.\cite{Alloul-etal-67} are reproduced in Fig.~\ref{fig-alloul}
together with the fit of the form (\ref{Sosc}).


8.  Everything in the above treatment of conventional SE applies 
to the situations, when the inhomogeneous broadening is absent, and 
the second pulse is not $\pi$ but has a different
``angle''. This, in particular, includes the so-called solid 
echoes in one-component 
spin systems (but excludes\cite{Fine-04} the special pulse sequences
designed to reverse the spin dynamics\cite{Rhim-etal-71,Pastawski-etal-00}.)

9. One can thus perform a conclusive experimental test 
of the present theory by studying solid echoes in CaF$_2$.
This well-characterized material exhibits the oscillatory 
FID asymptotics\cite{Engelsberg-etal-74}.
Such an experiment can verify not only 
the result of the theory (i.e. the echo envelope) 
but also the underlying picture, namely
it can measure the process $M(t)$ and check whether 
the long-time behavior
of $\tilde{F}(\tau,t)$ is characterized by the same time constants
as $F(t)$ independently of the type of the pulse applied at $t=\tau$. 

\begin{figure} \setlength{\unitlength}{0.1cm}
\begin{picture}(100, 48) 
{ 
\put(0, -2){ \epsfxsize= 3.1in \epsfbox{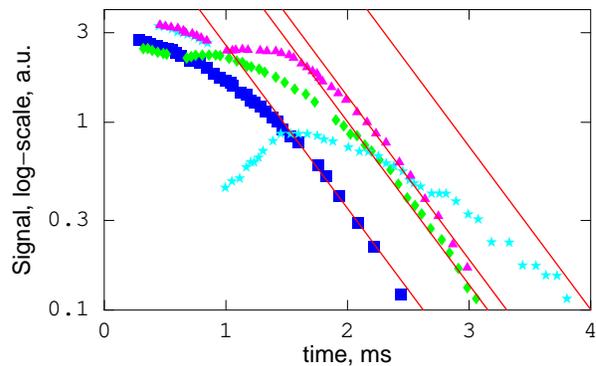} }
}
\end{picture} 
\caption{ 
Experimental data extracted from the oscilloscope photographs
of $^{129}$Xe transient signals presented in Fig.1 of 
Ref.\cite{Warren-etal-67}: squares --- FID; diamonds, triangles and stars
are the solid echoes with the pulse sequences (in respective order): 
$90^{\circ}-\tau -  90^{\circ}_{\ 90^{\circ}}$, 
$90^{\circ}-\tau -  180^{\circ}_{\ 0^{\circ}}$ and 
$90^{\circ}-\tau -  90^{\circ}_{\ 0^{\circ}}$. Solid lines are the 
exponential fits with the same time constant.
} 
\label{fig-warren} 
\end{figure}

10. The latter kind of test can, in fact, be performed on the old 
results reported for the solid mixture of 
$^{129}$Xe---$^{131}$Xe \cite{Warren-etal-67}. 
The experimental traces of $^{129}$Xe FID
and three different solid echoes are
reproduced in Fig.~\ref{fig-warren}. 
As seen in this figure, the long-time behavior
of the FID and that of two echoes
can be well fit by the same exponent.
The third echo, however, does not
approach clear exponential shape within the observational time window.
A conclusive test of the theory would, obviously, 
require more accurate measurements.

11. Echo studies\cite{Nakamura-etal-02} 
have indicated that some kind of slow fluctuators
of unknown microscopic origin are responsible
for decoherence in solid-state $q$-bits. It is possible
that, in each particular $q$-bit, a small number of such fluctuators
control the decoherence process. If true, the fluctuation rates
of those few fluctuators can be widely separated from each other
and, therefore, divided into "slow" and "fast" as shown in 
Fig.~\ref{fig-distrib}. In such a case, the present
theory is applicable and can be used to analyze the echo envelopes.
In this context, I would like to point out that the long-time behavior 
(\ref{Sosc}) is consistent with
the beats in the echo envelope observed in Ref.\cite{Nakamura-etal-02}.


In conclusion, the analysis presented in this work suggests that,
when the long-time behavior of SE is observed to be of the form (\ref{Smono})
or (\ref{Sosc}), the time constants of this decay can be extracted
from the corresponding homogeneous FID calculation. 
The experimental confirmation 
of the present theory (in particular, the solid echo 
experiment in CaF$_2$) would also give a strong 
support to the notion of the eigenvalues of time evolution operators
in quantum systems. 
The author is grateful to C.P.~Slichter, J.~Haase and M.~Mehring 
for discussing this work.

\bibliography{echo}

\end{document}